# Semiclassical approach to the description of the basic properties of nanoobjects

## Yuri Kornyushin

*Maître Jean Brunschvig Research Unit, Chalet Shalva, Randogne CH-3975, Switzerland*
E-mail: jacqie@bluewin.ch

Present paper is a review of results, obtained in the framework of semiclassical approach in nanophysics. Semiclassical description, based on Electrostatics and Thomas–Fermi model was applied to calculate dimensions of the electronic shell of a fullerene molecule and a carbon nanotube. This simplified approach yields surprisingly accurate results in some cases. Semiclassical approach provides rather good description of the dimensions of the electronic shell of a fullerene molecule. Two types of dipole oscillations in a fullerene molecule were considered and their frequencies were calculated. Similar calculations were performed for a carbon nanotube also. These results look rather reasonable. Three types of dipole oscillations in carbon nanotube were considered and their frequencies were calculated. Frequencies of the longitudinal collective oscillations of delocalized electrons in carbon peapod were calculated as well. Metallic cluster was modeled as a spherical ball. It was shown that metallic cluster is stable; its bulk modulus and the frequency of the dipole oscillation of the electronic shell relative to the ions were calculated.

PACS: **73.63.–b** Electronic transport in nanoscale materials and structures.

Keywords: fullerene molecule, carbon nanotube, metallic cluster.

## 1. Introduction

Semiclassical description, based on Electrostatics and Thomas–Fermi model was applied rather successfully to study atomic, molecular and nanoobjects problems [1–5]. Fullerene molecule was studied in [1,3–5]. Carbon nanotube was studied in [2,3–5]. Carbon peapod was studied in [5].

Fullerene molecule ($C_{60}$) forms a spherical ball. Let us denote the radius of the sphere, on which the carbon atoms are situated $R_f$. Chemical bonds determine the value of this radius. The electronic configuration of the constituent carbon atom is $1s^2 2s^2 2p^2$. It was assumed [1,3,4] that in a fullerene molecule two $1s$ electrons of each atom belong to the core (forming the ion itself), two $2s$ electrons form molecular bonds, and two $2p$ electrons are delocalized, or free. Hence it was assumed in [1,3,4] that the total number of the delocalized electrons in a fullerene molecule is 120. Another assumption was made, that the delocalized electrons couldn't penetrate inside the sphere of the ions as they repel each other.

The latest experimental data [6] show that $R_f = 0.354$ nm, the total number of the delocalized electrons, $N = 240$, the equilibrium internal radius of the electron shell, $R_{ie} = 0.279$ nm (that is the delocalized electrons do penetrate the sphere of the ions), and the equilibrium external radius of the electronic shell, $R_e = 0.429$ nm.

In the model considered here it is assumed like in [1,3–5] that the positive charge of the ions, $-eN(e < 0$ is the electron charge), is distributed homogeneously on the surface of a sphere $r = R_f$ ($r$ is the distance from the center of a fullerene molecule). The charge of the delocalized electrons, $eN$, is assumed to be distributed homogeneously in a spherical layer $R_i < r < R$ (see Fig. 1).

Now let us calculate the electrostatic energy of a fullerene molecule [5].

## 2. Calculation of the electrostatic energy of a fullerene molecule

The electrostatic potential $\varphi(\mathbf{r})$, arising from electric charge, can be obtained as a solution of the Poisson's equation [7]

$$\Delta \varphi(\mathbf{r}) = -4\pi \rho(\mathbf{r}), \qquad (1)$$

where $\rho(\mathbf{r})$ is the charge density.

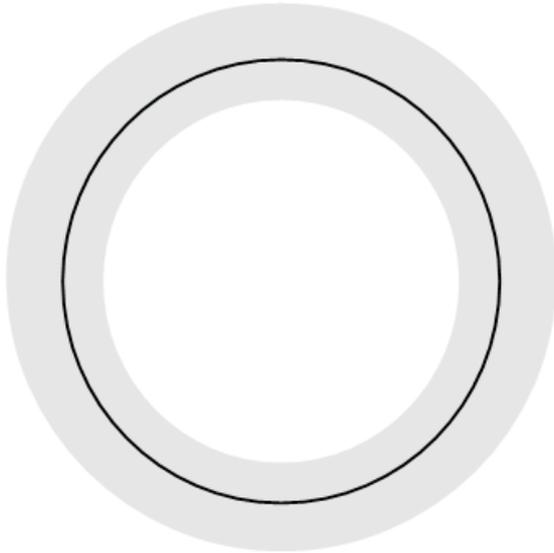

*Fig. 1.* Model, applied to a fullerene molecule and a carbon nanotube [5]. The charge of the ions is distributed homogeneously on the surface of a sphere of a radius $R_f$, or cylinder of a radius $R_n$ for a carbon nanotube. The delocalized electrons are distributed uniformly in a spherical (or cylindrical for a carbon nanotube) layer $R_i < r < R$.

The uniformly distributed charge density in the spherical layer of the electronic shell of a fullerene molecule is

$$\rho(\mathbf{r}) = 3eN/4\pi(R^3 - R_i^3) . \qquad (2)$$

The charge of the delocalized electrons produces the following electrostatic potential inside the spherical layer, where the delocalized electrons are found:

$$\varphi_e = [eN/(R^3 - R_i^3)][(3R^2/2) - (r^2/2) - (R_i^3/r)] . \quad (3)$$

The charge of the ions creates the following potential on the sphere $r = R_f$ and outside of it:

$$\varphi_i(r) = -eN/r . \qquad (4)$$

At $r \le R_f$ the positive charge of the sphere creates electrostatic potential equal to $-eN/R_f$. According to Eqs. (3), (4), the total electrostatic potential, $\varphi(r) = \varphi_e(r) + \varphi_i(r)$, is zero at $r = R$. Hence for $r > R$ it is zero also. Inside the neutral fullerene molecule, when $0 \le r \le R_i$, the total electrostatic potential is as follows:

$$\varphi_{in} = [3eN(R + R_i)/2(R^2 + RR_i + R_i^2)] - eN/R_f . \quad (5)$$

As on the external surface of a neutral fullerene molecule, at $r = R$, the total electrostatic potential is zero, it should be zero on the internal surface, at $r = R_i$, also. Otherwise the electrons will move to the locations, where their potential energy is smaller. In a state of equilibrium $\varphi_{in} = 0$. This yields:

$$R_i = 0.5\{R_f - \delta + [R_f^2 - (\delta^2/3)]^{1/2}\}, \qquad (6)$$

where $\delta = R - R_i$ is the thickness of the electronic shell. The latest experimental data [6] are: $R_f = 0.354$ nm and $\delta = 0.15$ nm. For these values Eq. (6) yields $R_i = 0.274$ nm. The experimental value is 0.279 nm. The calculated value is in very good agreement with the observed one.

The total electrostatic energy of a system, $U$, consists of the electrostatic energy of the ions, electrostatic energy of the delocalized electrons, and electrostatic energy of the interaction between them. It is described by the following expression:

$$U = (e^2N^2/2R_f) + (e^2N^2/2R) -$$
$$- [e^2N^2/2R_f(R^3 - R_i^3)](3R_fR^2 - 2R_i^3 - R_f^3) +$$
$$+ 0.1[eN/(R^3 - R_i^3)]^2 [R^5 + 9R_i^5 - 5(R_i^6/R) - 5R_i^3R^2] . \quad (7)$$

Here the first term describes the electrostatic energy of the ions. The third term describes the interaction energy. The second and the fourth term describe the electrostatic energy of the delocalized electrons. The second term represents the energy of the electrostatic field outside the charged spherical layer. The fourth term represents the energy of the field inside the spherical layer. This term was calculated as an integral over the layer of the square of the gradient of $\varphi_e(r)$, divided by $8\pi$.

## 3. Calculation of the kinetic and total energies of a fullerene molecule

The value of the kinetic energy, $T$, of a gas of delocalized electrons, confined in a volume of regarded spherical layer $4\pi(R^3 - R_i^3)/3$, and calculated in accord with the ideas of Thomas–Fermi model, is [1]

$$T = 1.105(\hbar^2/m)N^{5/3}/(R^3 - R_i^3)^{2/3} . \qquad (8)$$

The total energy, $W = T + U$, is a function of a single variational parameter, $\delta$. Indeed, when Eq. (6) is substituted in Eqs. (7), (8) together with $R = R_i + \delta = 0.5 \times \{R_f + \delta + [R_f^2 - (\delta^2/3)]^{1/2}\}$, we see that the total energy is a function of a single variational parameter, $\delta$. In a state of equilibrium the total energy reaches a minimum value $W_{min} = 9.422$ keV. The minimum (equilibrium) value of the total energy is reached when $\delta = \delta_e = 0.1374$ nm. At that $R_i = R_{ie} = 0.2808$ nm, and $R = R_e = 0.4182$ nm.

The minimum (equilibrium) energy of a positive ion of a fullerene molecule (with $N = 239$) was calculated to be 10.327 keV. At that in this case $R_{ie} = 0.2808$ nm and $R_e = 0.4182$. The difference, appearing in further digits, is very small.

It is worthwhile noting that the total energy of a fullerene molecule also contains large negative terms, corresponding to the energy of the ionic cores and the energy of the chemical bonds. These terms were not calculated here.

## 4. Collective dipole oscillation of delocalized electrons in a fullerene molecule

Linear-response theory was used by G.F. Bertsch et al. to calculate the frequency of a giant collective resonance in a fullerene molecule [8]. The value, calculated in [8] was of about 20 eV. This value, predicted in [8], was measured experimentally and reported in [9].

Two peaks of dipole collective oscillations of delocalized electrons in $C_{60}$ positive ions, observed experimentally by S.W.J. Scully et al., were reported in [10]. The authors associated the lower peak near 20 eV with a surface plasmon, its frequency being $\sqrt{3}$ times smaller than the Langmuir frequency [2]. The other peak (about 40 eV) is associated by the authors with some volume plasmon of unknown origin [10].

*Mie Oscillation.* Mie oscillation is a collective delocalized electron oscillation in a thin surface layer of a conducting sphere [11,12], where positive and negative charges are not separated in space. So it does not look suitable to regard such an oscillation in the thin surface layer of the electronic shell of a fullerene molecule. Anyway let us estimate possible frequency of such oscillation. Let us consider a thin surface layer of a thickness $d$, situated on the external surface of the electronic shell of a fullerene molecule. The volume of this layer, $4\pi R_e^2 d$, contains electric charge $4\pi R_e^2 end$ (here $n$ is the number of delocalized electrons per unit volume of the electronic shell). Let us shift the layer as a whole by a distance $s$ along the $z$-axis. This shift creates a dipole moment $P = 4\pi R_e^2 ensd$ and restoring electric field (in a spherically symmetric object) $E_r = 4\pi P/3 = (4\pi R_e)^2 ensd/3$ correspondingly. Restoring force at that is $(4\pi enR_e)^2 sd/3$. The mass of the regarded thin surface layer is $4\pi R_e^2 nmd$ (here $m$ denotes the mass of the delocalized electron). Restoring force leads to the dipole oscillation. From Newton equation follows that the frequency of the oscillation considered is $\omega_M = (4\pi e^2 n/3m)^{1/2}$. This frequency, as was mentioned by S.W.J. Scully et al. [10], is $\sqrt{3}$ times smaller than Langmuir frequency. In the model described above $n = 3N/4\pi(R_e^3 - R_{ie}^3)$. From two last equations written above follows that $\omega_M = [e^2 N/m(R_e^3 - R_{ie}^3)]^{1/2}$. This equation yields $\hbar\omega_M = 21.45$ eV when the experimental values of the parameters are used. S.W.J. Scully et al. reported the value of one of the two measured peaks, $\hbar\omega_M$, near 20 eV [10]. The agreement is quite reasonable. Using theoretical values of $R_e = 0.4182$ nm and $R_{ie} = 0.2808$ nm, calculated here, one can obtain the value of the peak 22.72 eV. This result is not too bad either.

*Simple Dipole Oscillation.* Let us consider a dipole type collective quantum oscillation of the electronic shell as a whole (having mass $mN$) relative to the ion skeleton. Such an oscillation looks more plausible in a fullerene molecule than Mie oscillation. Regarded system consists of two spherically symmetric objects, the ion skeleton and the electronic shell. Hence this oscillation causes no electric field in the blank interior of a fullerene molecule around its center (cavity). The value of the electrostatic potential inside the electronic shell is $\varphi_e = [eN/(R_e^3 - R_{ie}^3)][(3R_e^2/2) - (r^2/2) - (R_{ie}^3/r)]$ [see Eq. (3)]. Let us consider the shift of the electronic shell by a small distance, $\mathbf{s}$, relative to the ion skeleton and calculate the change in the electrostatic energy of a fullerene molecule (in the framework of the accepted model). The first term in Eq. (3) does not contribute to the change in the electrostatic energy, as it is constant. The contribution of the third term is also zero because the third term depends on the distance like $1/r$, a potential produced by a point charge. Interaction of the ion skeleton with the point charge, situated near the center of the fullerene molecule, does not depend on their relative position, because the electrostatic potential is constant for $r \leq R_f$. The change in the electrostatic energy does not depend on the direction of the shift ($\mathbf{s}$ vector) because of the spherical symmetry of the problem. This means that the change cannot contain term proportional to $s$. Let us choose the $x$-axis along the $\mathbf{s}$ vector. Then after the shift we shall have in the third term of Eq. (3) $[(x - s)^2 + y^2 + z^2]$ instead of $r^2 = (x^2 + y^2 + z^2)$. Taking into account that the term, proportional to $s$ in the change of the electrostatic energy is zero, we find that the third term (the only one, which contributes) yields the following change in the electrostatic energy of a fullerene molecule due to the shift regarded: $U(s) = (eNs)^2/2(R_e^3 - R_{ie}^3)$ [5]. This energy, $U(s)$, is a potential of a 3-$d$ harmonic oscillator [13]. For a 3-$d$ harmonic oscillator with mass $m NU(s) \equiv m N\omega_s^2 s^2/2$ [5,13]. As follows from the last two equations, the frequency, $\omega_s$, is equal to Mie frequency, $\omega_M$. The distance between the quantum levels is $\hbar\omega_M$. Its calculated value is 21.45 eV. Its experimental value is about 20 eV. Quantum excitation to the first excited level manifests itself as 20 eV peak, the second one as 40 eV peak. Conception of some volume plasmon of unknown origin [10] is not needed.

So, the frequency of the dipole oscillation in a fullerene molecule is as follows:

$$\omega_f = [e^2 N/m(R_e^3 - R_{ie}^3)]^{1/2}. \qquad (9)$$

Mie oscillation is also a 3-$d$ (quantum) harmonic oscillator. So the results are the same for both types of oscillations. Only Mie oscillation in a fullerene molecule looks less plausible.

The frequency of the simple dipole oscillation is equal to Mie frequency in spherically symmetric objects [1]. So it is rather difficult to determine which type of the oscillation occurs in the object under investigation.

## 5. Calculation of the electrostatic energy of a carbon nanotube

Let us consider a long (comparative to its diameter) carbon nanotube. Let us denote $N_n$ the number of delocalized electrons in a carbon nanotube per unit length. We denote the radius of the cylinder, on which the ions of a carbon nanotube are situated, $R_n$.

The Gauss theorem [7] allows calculating the electrostatic field inside the electronic shell of a long carbon nanotube (at $R_i \le r \le R$):

$$E_e(r) = (2eN_n/r)(r^2 - R_i^2)/(R^2 - R_i^2) . \quad (10)$$

The electrostatic field, produced by the ions at $R_n \le r \le R$ can be calculated also [7]:

$$E_i(r) = -2eN_n/r . \quad (11)$$

At $r \le R_n$ the electrostatic field, produced by the ions, is zero. The electrostatic potential difference between the internal and external surfaces of the electronic shell in a carbon nanotube should be zero in a state of equilibrium. Otherwise the delocalized electrons will move to the locations, where their potential energy is smaller. The electrostatic potential difference is an integral of minus electrostatic field between the surfaces mentioned. It is zero when

$$R_i = R^2/R_n \exp 0.5 . \quad (12)$$

Eq. (12) yields for the thickness of the electronic shell, $\delta = R - R_i$, the following result:

$$\delta = (R_i R_n)^{1/2}(\exp 0.25) - R_i . \quad (13)$$

As $\delta = R - R_i$, it follows from Eq. (13) that

$$R = R_i + \delta = (R_i R_n)^{1/2} \exp 0.25 . \quad (14)$$

The total electrostatic energy of a system is equal to the integral of the square of the electrostatic field, divided by $8\pi$. The value of it per unit length of a nanotube is described by the following expression:

$$U = [e^2 N_n^2/(R^2 - R_i^2)^2][R^4 \ln (R/R_n) + R_i^4 \ln (R_n/R_i)] + [e^2 N_n^2/(R^2 - R_i^2)](R_n^2 - 0.75R^2 - 0.75R_i^2) \quad (15)$$

Here [see Eq. (14)], $R = (R_i R_n)^{1/2} \exp 0.25$. From this follows that the electrostatic energy of the system is a function of only one variational parameter, $R_i$.

## 6. Calculation of the kinetic and total energies of a carbon nanotube

The kinetic energy per unit length, $T$, of a gas of delocalized electrons, confined in a cylindrical layer of a volume per unit length of a nanotube, $\pi(R^2 - R_i^2)$, and calculated in accord with the ideas of the Thomas–Fermi model, is as follows [2,3]:

$$T = 1.338 (\hbar^2/m)N_n^{5/3}/R_i^{2/3}[R_n(\exp 0.5) - R_i]^{2/3} . \quad (16)$$

Eq. (14) was used here to express the kinetic energy as a function of $R_n$ and one variational parameter, $R_i$. Thence, the total energy, $W = T + U$, is a function of a single variational parameter, $R_i$. In a state of equilibrium the total energy reaches a minimum value $W_{\min} = 71.51$ keV/nm at $R_n = 0.7$ nm and $N_n = 670$ nm$^{-1}$. Parameter $R_n = 0.7$ nm was taken from [4]. It was assumed here also that the total number of delocalized electrons in a carbon nanotube is four times larger than the number of carbon atoms. For the same values of $R_n = 0.7$ nm and $N_n = 670$ nm$^{-1}$ the minimum (equilibrium) value of total energy is reached when $R_i = R_{nie} = 0.577$ nm. At that $R = R_{ne} = 0.816$ nm and $\delta_e = 0.239$ nm.

It is worthwhile noting that the total energy of a carbon nanotube also contains large negative terms, corresponding to the energy of the ionic cores and the energy of the chemical bonds. These terms were not calculated here.

## 7. Collective dipole oscillations of delocalized electrons in a carbon nanotube

Let us consider first a longitudinal surface dipole collective oscillation of delocalized electrons in a long carbon nanotube [2]. The electric current can exist only in the layer occupied by collective electrons. In this layer the current carriers can move in an oscillatory fashion. We assume that the collective electrons move as a whole without deformation along the axis of a carbon nanotube relative to the ions. We shall call such an oscillation longitudinal. Let us shift the current carriers in the layer by a distance $h$ along the axis of a cylinder. As a result of the shift we have opposite charges on the bases of a cylinder, $q = \pm eN_n h$. The two bases are situated rather far away from each other, so they should be treated as two separate charged disks. A disk is a limiting case of a flattened ellipsoid of revolution with half-axes $a < b = c$. So the capacitance of a disk is that of an ellipsoid with $a = 0$ and $b = c = R_{ne}$, that is the capacitance, $C = 2R_{ne}/\pi$ [14]. The electrostatic energy of the two disks is $U_d = 2 \cdot 0.5q^2/C = (\pi e^2 N_n^2 / 2R_{ne})h^2$. The force, acting on all the electrons as a whole, is a derivative $-dU_d/ds = -\pi e^2(N_n^2/R_{ne})h$. Thus we come to the Newton equation for the $lN_n$ collective electrons as a whole:

$$mlN_n(d^2h / dr^2) + \pi e^2(N_n^2/R_{ne})h = 0 . \quad (17)$$

Assuming $h = h_0 \sin \omega_l t$, and using Eq. (17), we come immediately to the expression:

$$\omega_l = e(\pi N_n/mlR_{ne})^{1/2} . \quad (18)$$

The angular frequency of the longitudinal oscillation is inversely proportional to the square root of the length of the carbon nanotube [2]. At $N_n = 670$ nm$^{-1}$, $l = 30$ nm, $R_{ne} = 0.816$ nm we have $\hbar\omega_l = 3.065$ eV.

Surface dipole collective oscillation of delocalized electrons in a conductive sphere is called, as was mentioned above, Mie oscillation [11,12]. The frequency of Mie oscillation is square root from 3 times smaller than Langmuir frequency [2], $\omega_L = (4\pi e^2 n/m)^{1/2}$ (here $n$ is the number of the delocalized electrons per unit volume of the electronic shell). In general case [12] the frequency of Mie oscillation, $\omega_M = (4\pi De^2 n/m)^{1/2}$ (here $D$ is the depolarization factor [12]). Since for a sphere [7] $D = 1/3$, we have for the frequency of Mie oscillation in a sphere $\omega_M = (4\pi e^2 n/3m)^{1/2}$, as was mentioned above. For a cylinder [7] $D = 1/2$ and $n = N_n/\pi(R_{ne}^2 - R_{nie}^2)$. Thence we have for the frequency of the transverse surface (Mie) oscillation in carbon nanotube following result:

$$\omega_M = [2e^2 N_n/m(R_{ne}^2 - R_{nie}^2)]. \qquad (19)$$

For $N_n = 670$ nm$^{-1}$, $R_{ne} = 0.816$ nm, and $R_{nie} = 0.577$ nm we have $\hbar\omega_M = 21$ eV.

Let us consider now a simple dipole oscillation of the electronic shell as a whole relative to the ion skeleton perpendicular to the carbon nanotube axis. Regarded system consists of two cylindrically symmetric objects, the ion skeleton and the electronic shell. Hence this oscillation causes no electric field in the blank interior of a carbon nanotube around its center (cavity).

Like for a fullerene molecule, when the electronic shell of a carbon nanotube is shifted as a whole by a small distance, **h**, relative to the ion skeleton, the electrostatic energy of a carbon nanotube is changed by $W(h)$, which is a potential of a $2-d$ harmonic oscillator. Let us calculate this change. The electrostatic potential is equal to the integral of minus electric field [7]. Inside the electronic shell of a carbon nanotube, as follows from Eq. (10), the electrostatic potential is as follows:

$$\varphi_e = -[eN_{en}r^2/(R_{ne}^2 - R_{nie}^2)] - \\ - 2eN_n[R_{nie}^2/(R_{ne}^2 - R_{nie}^2)]/\ln(r/r_0), \qquad (20)$$

where $r_0$ is some constant. The second term in Eq. (20) coincides with the potential produced by some linear charge. This term does not contribute to the change in the electrostatic energy of a carbon nanotube, because the interaction between the ion skeleton and a linear charge is zero. This is so because the electrostatic potential is constant for $r \leq R_n$. The change in the electrostatic energy does not contain a term, proportional to $h$ because of the radial symmetry of the problem. So the only contribution comes from the first term of Eq. (20). Let us choose the $x$-axis along the **h** vector. Then after the shift we shall have in the first term of Eq. (20) $[(x-h)^2 + y^2]$ instead of $r^2 = (x^2 + y^2)$. Taking into account that the term, proportional to $h$ in the change of the electrostatic energy is zero, we find that the first term (the only one, which contributes) yields the following value of the change in the electrostatic energy of a carbon nanotube (per unit length) due to the shift regarded: $U(h) = (eN_n h)^2/(R_{ne}^2 - R_{nie}^2)$ [5]. This energy, $U(h)$, is a potential of a $2-d$ harmonic oscillator [5,13]. For a $2-d$ harmonic oscillator [13] (mass per unit length of a carbon nanotube $mN_n) U(s) \equiv \equiv mN_n\omega_s^2 s^2/2$. As follows from the last two equations, the frequency, $\omega_s$, is equal to Mie frequency, $\omega_M$. The distance between the levels is $\hbar\omega_M$. Its calculated value is 21 eV. So, according to the calculations performed, the quantum transition to the first excited level should manifest itself as 21 eV peak, and to the second one as 42 eV peak.

Mie oscillation in a carbon nanotube is also a $2-d$ harmonic oscillator. So the results are the same for both types of the oscillations. Only Mie oscillation in a carbon nanotube looks less plausible.

Obtained results show that a very simple semiclassical concept of Thomas–Fermi model and Electrostatics often gives rather good agreement between experimental and theoretical results. In particular, this model often gives rather reasonable description of nanoobjects.

## 8. Carbon peapod

As in [15] we assume that the fullerene molecules are encapsulated in a carbon nanotube, and there is an interaction between collective electrons of a carbon nanotube and fullerene molecules.

Let us consider the longitudinal oscillations [4]. We remind that $R_{ne}$ denotes equilibrium external radius of the electronic shell of a carbon nanotube and $R_{nie}$ denotes equilibrium internal radius of a carbon nanotube. Let us calculate the part of the electrostatic energy (per unit length), which depends on the shift of the electronic shell of a fullerene molecule $s$ and that of a carbon nanotube $h$. It consists of the energy of the charges arising on the ends of a carbon nanotube, $2 \cdot 0.5 q^2/lC = (\pi e^2 N_n^2/2lR_{ne})h^2$, the energy of a fullerene molecule, $U(s) = (eNs)^2/2(R_e^3 - R_{ie}^3)$, multiplied by the number of fullerene molecules in a unit length of a carbon peapod $n_f$, the energy of the interaction of the dipole moments of the fullerene molecules $en_f Ns$ with electric field inside a carbon nanotube $E_n = (2q/lC) = = \pi e(N_n/lR_{ne})h$ [this interaction energy is equal to $\pi e^2(n_f NN_n/lR_{ne})sh$], and the energy of a dipole–dipole interaction between the dipoles of fullerene molecules. The last term was calculated for a linear array of fullerene molecules with the assumption that the size of the dipoles is smaller than the distance between them. The part of the total electrostatic energy, depending on the shifts $s$ and $h$ is as follows:

$$U(s,h) = e^2 n_f N^2 \{[0.5/(R_e^3 - R_{ie}^3)] - $$
$$ - 2(n_f^2/l)f(n_f l)\}s^2 + (\pi e^2 N_n^2/2lR_{ne})h^2 \pm$$
$$ \pm \pi e^2 (n_f NN_n/lR_{ne})sh,$$

$$\text{where} \quad f(k) = k \sum_{i=1}^{k-1}(1/i^3) - \sum_{i=1}^{k-1}(1/i^2). \quad (21)$$

The term $-2e^2N^2(n_f^3/l)f(n_f l)s^2$ in Eq. (21) represents the energy of the dipole–dipole interaction in a periodic array of fullerene molecules, calculated in the approximation which neglects the real size of a dipole. This approximation somewhat underestimates the effect of dipole–dipole interaction.

The last term in Eq. (21) can be positive or negative depending on the direction of the shifts $s$ and $h$. It is positive when the shifts are parallel, and negative when they are antiparallel.

Let us consider now the transverse oscillations. In accepted here model the shift of the electronic shell (as a whole) relative to the ion skeleton does not produce electrostatic field inside the fullerene molecule or carbon nanotube. This is because the ion skeleton and the electronic shell do not lose their symmetry as a result of the shift. So there is no interaction between transverse oscillations of a fullerene molecule and carbon nanotube. This means that the two transverse oscillations are not changed in a carbon peapod; they remain the same as they were in a free fullerene molecule and a free carbon nanotube.

Let us write the energy of the longitudinal oscillation $U(s,h)$ in the following form:

$$U(s,h) = 0.5As^2 + 0.5Bh^2 + Gsh. \quad (22)$$

The specific values of the factors $A$, $B$, and $G$ are given in Eq. (21).

Now let us write Newton equations for all the collective electrons per unit length of a carbon nanotube, and for all the collective electrons of all the fullerene molecules per unit length of a carbon nanotube, for the case of the longitudinal oscillations:

$$mN_n(\partial^2 h/\partial t^2) + Bh \pm Gs = 0,$$
$$mn_f N_f(\partial^2 s/\partial t^2) + As \pm Gh = 0. \quad (23)$$

Two Eqs. (23) yield the following equations for the amplitudes of the oscillations $h_0$ and $s_0$:

$$(B - mN_n\omega^2)h_0 \pm Gs_0 = 0,$$
$$\pm Gh_0 + (A - mn_f N\omega^2)s_0 = 0. \quad (24)$$

Eqs. (24) yield the frequencies of the oscillations (the same frequencies for both parallel and antiparallel modes):

$$\omega_{1,2}^2 = 0.5\{(\omega_f^2 - 2\omega_a^2 + \omega_l^2) \pm$$
$$\pm [(\omega_f^2 - 2\omega_a^2 - \omega_l^2)^2 + 4\pi n_f(R^3/lR_{ne})\omega_f^2\omega_l^2]^{1/2}\},$$
$$(25)$$

$$\omega_a^2 = 2e^2 n_f^2 N_f f(n_f l)/lm. \quad (26)$$

Here $\omega_f$ is the angular frequency of the plasmon in an isolated $C_{60}$ molecule [Eq. (9)], $\omega_l$ is the angular frequency of the longitudinal oscillations in an isolated carbon nanotube [Eq. (18)]; $(\omega_f^2 - 2\omega_a^2)^{1/2}$ is the angular frequency of the dipole mode of the plasma longitudinal oscillations of a linear periodic array of fullerene molecules. We find that the dipole–dipole interaction between the fullerene molecules reduces the plasmon energy for longitudinal oscillations. We also find in the accepted model that in the limit of infinite length carbon peapod, the coupling between the carbon nanotube and $C_{60}$ molecules vanishes.

Eqs. (25), (26) do not take into account spatial dispersion (dependence on the wave vector $\mathbf{q}$). In [4] Eqs. (25), (26) were generalized, taking into account spatial dispersion for small wave vectors. Figure 2 shows the coupled longitudinal modes in a finite length carbon peapod (Fig. 1 in [4]), where the dotted line presents $\omega_l$. For small wave vector $\mathbf{q}$, the coupling between the carbon nanotube and $C_{60}$ molecules has a negligible effect. In the vicinity of $\omega_l = (\omega_f^2 - 2\omega_a^2)^{1/2}$, the coupling lowers the carbon nanotube branch while it raises the fullerene branch, forming an anticrossing.

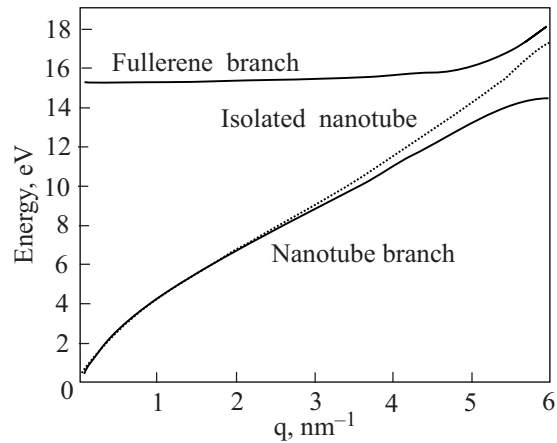

*Fig. 2.* Plasmon energy as a function of wavevector $\mathbf{q}$ along the carbon nanotube direction, for a carbon peapod of finite length (as calculated in [4]). The dotted curve is the plasmon dispersion of an isolated carbon nanotube, from Fig. 2 of [16]. The lower (higher) solid curve is the carbon nanotube (fullerene) branch, when coupling between the carbon nanotube and $C_{60}$ molecules is taken into account. Here the following parameters values were used [4]: $\hbar\omega_f = 20$ eV, $N = 120$, $n_f = 1.03$ nm$^{-1}$, $N_n = 333$ nm$^{-1}$, $R_e = 0.538$ nm, $R_f = R_{ie} = 0.353$ nm, $R_{ne} = 0.885$ nm, $R_n = R_{nie} = 0.7$ nm, and $l = 29.1$ nm ($n_f l = 30$). With these values, we have $f(n_f l) = 34.4$, $\hbar\omega_a = 9.07$ eV, $4\pi(n_f R_e^3/lR_{ne}) = 0.0783$, and $\hbar(\omega_f^2 - 2\omega_a^2)^{1/2} = 15.3$ eV.

## 9. Metallic cluster

A double-jelly model (delocalized electrons jelly and ions jelly) was applied for the description of metallic clusters [1]. This model does not take into account microscopic electrostatic field around the ions. Here we shall take into account this microscopic field. We shall analyze here the degenerate delocalized electrons.

A spherical ball models the shape of a cluster. We shall model the conditions as adiabatic ones under given pressure. We shall restrict our consideration here by given entropy and given pressure condition. In this case the thermodynamic potential, having a minimum in the state of equilibrium, is the enthalpy (it is a function of thermodynamic variables, entropy $S$ and pressure $P$: $H(S, P)$ [17]). Entropy is a function of a radius of a cluster $R$. So $R$ could be taken as thermodynamic variable instead of entropy. Under different condition the contribution of the entropy term to the Gibbs free energy is negligible.

We shall consider here only atmospheric pressure, which can be neglected. So really, we will take into account the energy of a cluster only.

### 9.1. Electrostatic energy of a separate ion

Let us consider a ball of a metallic cluster of a volume $V = 4\pi R^3/3$ (here $R$ is the radius of a cluster), consisting of $N_c$ delocalized electrons. We consider here the ions as point charges, and the delocalized electrons like a negatively charged gas. In metallic clusters the delocalized electrons density is so high, that they are degenerate over all the temperature range of the existence of a metallic cluster. Degenerate delocalized electrons screen a long-range electrostatic field of point charges. The screening Thomas–Fermi radius is as follows [18]:

$$1/g = (VE_F/6\pi N_c e^2)^{1/2} = -0.4714 R^{3/2} (E_F/N_c)^{1/2}/e,$$
$$(27)$$

where $E_F = (3\pi^2 N_c/V)^{2/3}(\hbar^2/2m)$ is the Fermi energy [18].

Fermi energy is proportional to $1/R^2$. Thomas–Fermi radius is proportional to $\sqrt{R}$.

The electrostatic field around a separate positive ion, submerged into the gas of degenerate delocalized electrons, is as follows [18]:

$$\varphi = -(ze/r)\exp(-gr),$$
$$(28)$$

where $z$ is the number of delocalized electrons per one atom, an $r$ is the distance from the center of the ion.

The electrostatic energy of this field is the integral over the ball volume of its gradient in a second power, divided by $8\pi$. The lower limit of the integral on $r$ should be taken as $r_0$ (the radius of the ion), a very small value. Otherwise the integral diverges. Calculation yields the following expression for the electrostatic energy of a separate ion:

$$U_0 = 0.5z^2 e^2 (r_0^{-1} + 0.5g)\exp{-2gr_0}.$$
$$(29)$$

For $2gr_0$ essentially smaller than unity Eq. (29) yields:

$$U_0 = (z^2 e^2/2r_0) - 0.75z^2 e^2 g.$$
$$(30)$$

The first term in the right-hand part of Eq. (30), $z^2 e^2/2r_0$, represents the electrostatic energy of the bare ion.

It is worthwhile to note that the expansion of a cluster leads to the decrease in the delocalized electron density. From this follows the increase in the screening radius [see Eq. (27)]. The electrostatic energy of a separate ion increases concomitantly. One can see it, regarding Eq. (30).

### 9.2. Total energy of a cluster

We regard the ions of a cluster as randomly distributed. It is well known since 1967, that the electrostatic energy of $N_c/z$ randomly distributed ions is just $U = (N_c/z)U_0$ [19].

The kinetic energy, $T$, of a gas of delocalized electrons, confined in a volume $(4/3)\pi R^3$, and calculated in accord with the ideas of the Thomas–Fermi model, is as follows [1]:

$$T = 1.105 N_c^{5/3}(\hbar^2/mR^2).$$
$$(31)$$

So the total energy of a cluster, $W(R)$, is as follows:

$$W(R) = 1.105 N_c^{5/3}(\hbar^2/mR^2) + (ze^2 N_c/2r_0) - 0.75ze^2 N_c g.$$
$$(32)$$

As a function of $R$, $W(R)$ has a minimum. So, the cluster is stable. The minimum value of $W(R)$ is

$$W_e = (ze^2 N_c/2r_0) - 0.565z^{4/3}N_c(me^4/\hbar^2).$$
$$(33)$$

This minimum occurs when $R = R_e$:

$$R_e = 2.422(N_c/z^2)^{1/3}(\hbar^2/me^2).$$
$$(34)$$

The equilibrium volume per one ion is

$$v_e = 4\pi z R_e^3/3N_c = 59.532(\hbar^2/me^2)^3/z.$$
$$(35)$$

For $z = 1$ $v_e = 8.821 \cdot 10^{-24}$ cm$^3$. It is about 3 times smaller than expected. For this value of $v_e$ we have average interatomic distance $2.564 \cdot 10^{-8}$ cm and $2/g = 5.27 \cdot 10^{-9}$ cm, which is 2.43 times smaller than average interatomic distance. This means that electrostatic fields of adjacent ions [Eq. (28)] do not overlap. So the electrostatic energy of $N_c/z$ ions is $U = (N_c/z)U_0$ anyway, are they distributed randomly or not. Overlapping takes place when $v_e \leq 0.2871(\hbar^2/me^2)^3/z$ only. According to Eq. (35) this is not so at ambient pressure.

### 9.3. Bulk modulus of a cluster

When condensed matter subject is expanded, the increase in its elastic energy is as follows [20]:

$$\delta W = 0.5K(\delta V)^2/V_e = 4.5KV_e(\delta R/R_e)^2 , \quad (36)$$

where $K$ is the bulk modulus, $V_e$ is the initial equilibrium volume, and $R_e$ is the initial equilibrium radius. Here a well-known relation, $\delta V/V_e = 3\delta R/R_e$, was used. It is valid when $\delta R$ is essentially smaller than $R_e$.

The change in the total energy of a cluster [Eq. (32)] is

$$\delta W = 0.5(\partial^2 W/\partial R^2)_{R=R_e}(dR)^2 . \quad (37)$$

Equations (36), (37) yield:

$$K = (1/12)\pi R_e (\partial^2 W/\partial R^2)_{R=R_e} . \quad (38)$$

Equations (32), 34), and (38) yield the following expression for the bulk modulus:

$$K = 0.00105(z^{10/3}m^4e^{10}/\hbar^8) . \quad (39)$$

For $z = 1$ Eq. (13) yields $K = 3.102 \cdot 10^{11}$ erg/cm$^3$. This value is a very reasonable one indeed.

### 9.4. Collective oscillations

Frequency of the dipole oscillation of delocalized electrons, $\omega_d$, in any conductor of a spherical shape is $\sqrt{3}$ times smaller than Langmuir plasma frequency [12]. For spherical cluster regarded here we have

$$\omega_d = 0.2653z^{1/2}me^4/\hbar^3. \quad (40)$$

For $z = 1$ we have $\hbar\omega_d = 7.219$ eV. This value is close to the classical surface-plasmon frequency in the case of Na [21]. For Na instead of the factor 0.2653 in Eq. (40) there is a factor 0.2497 [21]. So for Na we have $\hbar\omega_d = 6.795$ eV. This value is only 5.87% less then Eq. (40) yields. Nobody could expect that some simple approach like that used in this paper could compete with modern techniques of Theoretical Physics used in [21,22]. Semiclassical approach works only for simple basic problems. So far it works fine for the problems discussed here. But it cannot describe like in [22] a wide range of photon energies. It was shown in [22] that for lower energies collective effects are essential and for higher energies (higher than 15 eV) a single-particle picture is relevant.

We are not discussing here the quadrupole and breathing modes, but it is worthwhile mentioning that both of them have an atomic branch of slow oscillations and an electronic one of fast oscillations [1].

### 10. Discussion

Today many authors use the model of the electronic shell accepted here. This model allows understanding basic experimental results. As a matter of fact this concept is present in papers [6,10]. Two peaks of dipole oscillations of the delocalized electrons in $C_{60}$ ions, observed experimentally by S.W.J. Scully et al., were reported in [10]. The authors ascribe these peaks to the surface (Mie) oscillation [11,12] and some bulk dipole mode. The energy values of the quanta of the two peaks were measured to be about 20 and 40 eV. Using experimental data on $N$, $R$ and $R_i$ and equation for the Mie frequency [2,10], one could calculate the energy of the quantum of the Mie oscillation, 21.45 eV.

Simple dipole oscillation of the electronic shell as a whole has the same frequency as Mie oscillation. Oscillation of the electronic shell as a whole relative to the skeleton of the ions is an oscillation of a $3-d$ (for a fullerene molecule) and $2-d$ (for a carbon nanotube) harmonic oscillator. The same is for Mie oscillation. As the frequencies of the two types of oscillation are the same, cited experimental data could be explained both ways.

Concerning carbon nanotubes more sophisticated objects are now under investigation. In [23] collective excitations in a linear periodic array of parallel cylindrical carbon nanotubes, consisting of coaxial cylindrical tubules, was studied. Multiwall carbon nanostructures were studies in [24].

In [25] dynamic screening of an atom, confined within a finite width fullerene molecule is a subject of research. The authors used a classical dielectric approach. Now let us consider a problem of possibility of the atom confinement inside a fullerene molecule. There is no classical reason to find neutral atom confined in the cavity of a fullerene molecule or a neutral carbon nanotube. There is no electric field there and electrostatic potential there is zero like at the infinite distance from a fullerene molecule or a carbon nanotube. It was discussed in Sections 4 and 7 of the present paper.

In this problem we have two objects, a fullerene molecule and an atom. Let us denote the ionization energy of one object as $I$ and the affinity of the other as $-A$. Let us transfer one electron with charge $e$ from one object to another one. The extra charge $\pm e$ is situated on the external surface of the electronic shell of a fullerene molecule, $r = R_e$. Atom (ion) is charged with the extra charge of the same value and opposite sign. The electrostatic potential on the external surface of the electronic shell of a fullerene molecule is equal to that in the cavity. It was discussed in Sec. 2 of this paper. Both of them are equal to $\pm e/R_e$. The electrostatic energy of the interaction of the two charges, one in the cavity and the other one on the surface of the electronic shell, is $-e^2/R_e$. The total energy is changed by $I - A - e^2/R_e$ as a result of a charge transfer.

When this quantity is negative and minimal, the confinement discussed is possible. Either positive or negative ion could be found inside a fullerene molecule. It depends for which ion the total change in the energy is minimal.

Using Eq. (3) it is possible to see that in case of a positive ion its energy in the cavity around the center of a fullerene molecule is smaller than that inside the electronic shell. So a positive ion occupies the cavity of a fullerene molecule. Positive ion is not localized in the center of a fullerene molecule, like it was assumed in [25]. It is expected to be found in any place in the cavity.

It is opposite for a negative ion. As follows from Eq. (3), the minimal value of its energy is achieved when a negative ion is situated on the surface, where the positive ions of a fullerene molecule are located, $r = R_f$. This energy is smaller than $I - A - e^2/R_e$, because the positive ion skelton attracts a negative ion. This attraction is not taken into account in the expression $I - A - e^2/R_e$.

In [26] a model of non-homogeneous oscillation of the shell of a fullerene molecule and a carbon nanotube was proposed. This model also can explain experimental results, obtained by S.W.J. Scully et al. [10]. But this model does not look inherent to the topic. That is why it was not discussed here.

We modeled the shape of metallic cluster here as a spherical ball. The delocalized electrons in metallic cluster are degenerate. Their kinetic energy is calculated in the spirit of Thomas–Fermi model. The delocalized electrons screen the electrostatic field of the ions. This field was calculated; it was shown that it contributes essentially to the energy of a cluster and its stability.

The equilibrium values of the energy of a cluster and its volume were calculated. The bulk modulus of a cluster and simple collective dipole oscillation of the electronic shell as a whole relative to the ion skeleton were calculated also.

Obtained results show that a very simple semiclassical concept, based on Electrostatics and Thomas–Fermi model often gives rather good agreement between experimental and theoretical results. In particular, this model gives rather reasonable description of nanoobjects.